\documentclass[11pt]{article}
\usepackage{amsfonts,amsmath,amssymb,bm,graphicx}
\makeatletter
\newcommand\figcaption{\def\@captype{figure}\caption}
\newcommand\tabcaption{\def\@captype{table}\caption}

\makeatother \textheight 24 true cm \textwidth 16.5 true cm
\title{ The NLO corrections of $H_{TC} \Pi ^0$and $\Pi ^ + \Pi ^ - $
pair production at the ILC in the TC2 model}
\author{{Qing-Peng Qiao$^1$, ~Zuo Li$^1$, ~Xue-Qian Li$^1$, and ~Xue-Lei Wang$^2$} \\
{\small \it 1. Department of Physics, Nankai University, TianJin
300071, China} \\{\small \it 2. Department of Physics, Henan Normal
University, Xinxiang, China}}
\date{}
\begin{document}
\maketitle {\bf Abstract}\\

As well known, if the Higgs boson were not observed at LHC, the
technicolor model would be the most favorable candidate responsible
for the symmetry breaking. To overcome some defects in the previous
model, some extended versions have been proposed. In the TC2 model
typical signature is existence of heavy $H_{TC}$ and technipion
$\Pi$. A direct proof of validity of the model is to produce them at
accelerator. Thus we study  the production rates of $ e^ + e^ - \to
H_{TC} \Pi ^0 $ and $e^ + e^ - \to \Pi ^ + \Pi ^ -$ at ILC  in the
topcolor-assisted technicolor (TC2) model. In fact, there is a flood
of models belonging to new physics which can result in products with
characteristics similar to $H_{TC}+\Pi$ of the TC2 model. Therefore
to distinguish this model from others one may need to investigate
some details by calculating the cross section to NLO. We indeed find
that the NLO corrections are significant, namely the ratio $\delta
\equiv (\sigma _{NLO} - \sigma _{LO} )/\sigma _{LO} $ in $ e^ + e^ -
\to H_{TC} \Pi ^0 $ exceeds 100\%
within a plausible parameter space. \\

{\bf Keywords  }TC2, top-pion, top-higgs, LOOPTOOLS\\

\section{Introduction} \hspace{0mm}\vspace{2mm}

The success of the standard model (SM) is not doubtful at all. On
the other aspect, however, the mechanism which breaks the
electroweak symmetry is not yet quite understood. In the typical
spontaneous symmetry breaking scheme the Higgs boson is required but
it so far evaded observation. In addition, there exist the prominent
problems of triviality and unnaturalness in the Higgs sector. Thus
alternative dynamical symmetry breaking schemes were proposed, among
the models, the technicolor model (TC) is the most favorable one
which was proposed  by Weinberg and Susskind
\cite{Susskind:1978ms,Weinberg:1979bn} independently.\\

The advantage of dynamical electroweak symmetry breaking (EWSB) is
that there the elementary scalar field is not introduced to be
responsible for the breaking, therefore, it can avoid the troubles
of triviality and unnaturalness. However, the initial TC model is
the simplest version and exposes some obvious defects. To remedy
those defects, several modified version have appeared
\cite{Dimopoulos:1979es,Eichten:1979ah} later. In order to explain
the large mass difference between the top quark and the bottom
quark, the topcolor-assisted technicolor (TC2) model was proposed by
Hill \cite{Hill:1994hp,Lane:1995gw,Lane:1998qi} to improve the
original one. Namely the TC2 model can naturally produce large top
quark mass and realize dynamical electroweak symmetry breaking.
Concretely, in this model, the top-color interaction makes a small
contribution to the EWSB, but indeed is responsible for the main
part of the top quark mass as $(1-\epsilon_t)m_{t}$ where
$\epsilon_t$ is a model-dependent parameter within a range of
$0.03<\epsilon_t<0.1$ \cite{Hill:2002ap}, whereas the TC interaction
plays the main role for breaking the electroweak gauge symmetry. The
extended TC (ETC) interaction gives rise to the masses of the
ordinary fermions (quarks and leptons) and a small portion
$\epsilon_t M_{t}$  of the top mass. One of the most general
characteristics of the TC2 model is  existence of three
isospin-triplet pseudo-Goldstone bosons called as top-pions ( $\Pi ^
\pm, \Pi ^ 0$) and one isospin-singlet boson$-$the top-Higgs
($H_{TC}$). Obviously, such new particles do not exist in the SM,
and their appearance can be treated as clear and definite signatures
of the new physics beyond the SM. To be consistent with the SM
phenomenology, the energy scale of the model must be sufficiently
high, say at TeV order, so that one needs to look for direct
production of such new
particles at high energy experiments.\\

Definitely, the LHC would be the first place to carry out such
exploration, but since at the hadron colliders, the background is
very complicated and it is hard to identify the signal. Instead, in
the ILC experiment which will be be running in the future, the
situation is much better. Starting with a relatively simple
situation, therefore in this work, we study a favorable channel for
the electron-positron collisions and will carry out some rigorous
calculation for the LHC case in our next work. Concretely, we
consider the production process $ e^ + e^ - \to
H_{TC} \Pi ^0 $ and  $e^ + e^ - \to \Pi ^ + \Pi ^ -$.\\

In our earlier work \cite{Wang:2005ixa}, the tree level contribution
was considered and one noticed that such processes may be observable
for the designed luminosity of ILC. On other aspect, there is a
flood of new physics models which also result in similar production
processes (with different new particles). To distinguish the TC2
model from others, some details about the production cross sections
and differential cross sections are needed. At the tree level, some
parameters are fed in by hand and only the order of magnitude is
estimated as long as the NLO is significant, so that  one cannot
tell the difference of various models, thus the NLO calculation may
become necessary. Therefore, in this paper, we carry out the
calculation to NLO and we find  that the NLO contribution is
significant and moreover, NLO corrections are quite different for $
e^ + e^ - \to
H_{TC} \Pi ^0 $ and  $e^ + e^ - -> \to \Pi ^ + \Pi ^ -$.\\

This paper is organized as follows. In Section 2  we will present
the theoretical formulation of the production rates for processes $
e^ + e^ - \to H_{TC} \Pi ^0 $ and  $e^ + e^ - \to \Pi ^ + \Pi ^ -$.
By inputting the model parameters, we obtain the numerical resultsa
in Section 3. Our conclusion and some discussions are drawn in the
last section.

\section{Theoretical Formulation}\hspace{0mm}\vspace{2mm}

In this section, we will present the theoretical formulation of the
cross sections for two processes $ e^ + e^ - \to H_{TC} \Pi ^0$ and
$ e^ + e^ - \to \Pi^{+} \Pi ^- $ up to NLO in the TC2 model.\\

\subsection{ For $ e^ + e^ - \to H_{TC}
\Pi ^0 $}\vspace{5mm}

In the TC2 model there are three relatively light physical top-pions
( $\Pi ^ \pm, \Pi ^ 0$) whose  couplings  to $t$ and $b$ quarks are
\cite{Burdman:1999sr,He:1998ie,He:2002fd}:
\begin{eqnarray}
\frac{m_t \tan \beta }{v}[iK_{UR}^{tt} K_{UL}^{tt*} \bar {t}_L t_R
\Pi ^0 + \sqrt 2 K_{UR}^{tt} K_{DL}^{bb\ast} \bar {b}_L t_R \Pi ^ -
+ h.c.],
\end{eqnarray}\\
where $\tan\beta =\sqrt {(v/v_\pi)^2- 1} $ and the top-pion decay
constant $v_{\pi}\simeq O(60-100)$ GeV\cite{He:1998ie,He:2002fd}. $v
= \sqrt 2 v_w \approx 246$ GeV is the electro-weak symmetry breaking
scale. $K_{UL}^{ij}$ are matrix elements of the unitary matrix
$K_{UL}$ from which the Cabibbo-Kabayashi-Maskawa (CKM) matrix $V$
can be derived as $V=K_{UL}^{-1}K_{DL}$ and the matrices $K_{UL}$
and $K_{UL}$ are responsible for transforming the weak-engenstates
into the mass-eigenstates of left-handed U-type and D-type quarks
respectively. $K_{UR}^{ij}$ are the matrix elements of the
corresponding right-handed rotation matrix $K_{UR}$. Their values
can be found in Ref. \cite{He:1998ie,He:2002fd}:
\begin{eqnarray}
K_{UL}^{tt}=K_{DL}^{bb}=1,\; \qquad K_{UR}^{tt}=1-\epsilon_t.
\end{eqnarray}\\
Here, there is a free parameter $0.03<\epsilon_t<0.1$ which was
discussed in the relevant literature about how the heavy top quark
and other light quarks obtain their masses from different sources
\cite{Hill:2002ap}.\\

The TC2 model also suggests existence of  a scalar $H_{TC}$ called
as the top-Higgs boson \cite{Burdman:1999sr,Leibovich:2001ev}, which
is a $\overline{t}t$ bound state and analogous to the $\sigma$ boson
which plays an important role for low energy phenomenology. Its
couplings to quarks are in analog to that of the neutral top-pions.
The Feynman rules related to the top-pions and the top-Higgs are
shown below \cite{Leibovich:2001ev}:

\begin{eqnarray}
\begin{array}{l}
 Z_\mu H_{TC} \Pi ^0:\frac{\displaystyle g}{\displaystyle 2c_w }\frac{\displaystyle v_{T} }
{\displaystyle v}(P_\mu ^H - P_\mu ^0 ),~~~~
 Z_\mu \Pi ^ - \Pi ^ + :i\frac{\displaystyle g}{\displaystyle c_w }(1 - 2s_w^2 )(P_\mu ^- - P_\mu ^+ ),
\\\\
 A_\mu \Pi ^ - \Pi ^ + :ie(P_\mu ^ - - P_\mu ^ + ), ~~~~~~~~~~~~
 \bar {t}t\Pi ^0:\frac{\displaystyle  -m_t \tan \beta }{\displaystyle v}(1 - \epsilon_t )\gamma _5 ,
\\\\
 t\bar {b}\Pi ^ + :i\sqrt 2 \frac{\displaystyle m_t \tan \beta }{\displaystyle v}(1 - \epsilon_t )L,
~~~~~~
 \bar {t}b\Pi ^ - :i\sqrt 2 \frac{\displaystyle m_t \tan \beta }{\displaystyle v}(1 - \epsilon_t )R,
 \\\\
 t\bar{t}H_{TC} :i\frac{\displaystyle m_t \tan \beta }{\displaystyle v}(1 - \epsilon_t ), \\\\
 \end{array}
\end{eqnarray}
where  $v^2 = v_\pi ^2 + v_T^2$ and $v_T$ is the techni-pion decay
constant similar to that of regular pions, $L=(1-\gamma_5)/2,
~R=(1+\gamma_5)/2,~s_{w}=sin\theta_{w}$ and $c_{w}=cos\theta_{w}$
($\theta_w$ is the Weinberg angle).\\

With these interaction vertices, we can immediately write down the
production amplitude of $ e^ + e^ - \to H_{TC}\Pi ^0 $ at the tree
level:

\begin{eqnarray}
M^0 = \frac{g^2}{2c_w^2 }\frac{v_T }{v}\bar {v}(P_1^{' })(\not
{P_2} - \not {P}_3 )[ - \frac{1}{4}(1 - \gamma _5 ) + s_w^2
]u(P_2^{'} )\frac{1}{P_1^2 - M_Z^2 }
\end{eqnarray}\\

At the NLO level, The Feynman diagrams responsible for the process
are shown in Fig. \ref{ZHP}. When carrying out the loop integration,
an ultraviolet (UV) divergences appears and one needs to renormalize
the Z-t-t coupling to remove the UV divergence. In this work, we
employ the modified renormalization scheme $\overline{MS}$. \\

The loop-induced amplitude is written as
\begin{eqnarray}
M^{loop} = \frac{g^2}{2c_w^2 }N\bar {v}(P_1^{'} )[f_2\not{{P_2}} +
f_3\not{{P}}_3 ][ - \frac{1}{4}(1 - \gamma _5 ) + s_w^2 ]u(P_2^{'}
)\frac{1}{P_1^2 - M_z^2 },
\end{eqnarray}
where N is the regular color factor. Calculations of such loop
diagrams are straightforward. Each loop integration is composed of
some scalar loop functions \cite{'t Hooft:1978xw}, which are
evaluated in terms of the code LOOPTOOLS
\cite{Hahn:1998yk,Hahn:2004rf}. The explicit expressions of relevant
form factors ($f_2$, $f_3$) are lengthy, so that we keep them in
Appendix A. The NLO amplitude is then written as
\begin{equation}
M=M^{tree}+M^{loop}.
\end{equation}
\\[\intextsep]
\begin{minipage}{\textwidth}
\centering
\includegraphics[height=3cm,angle=0]{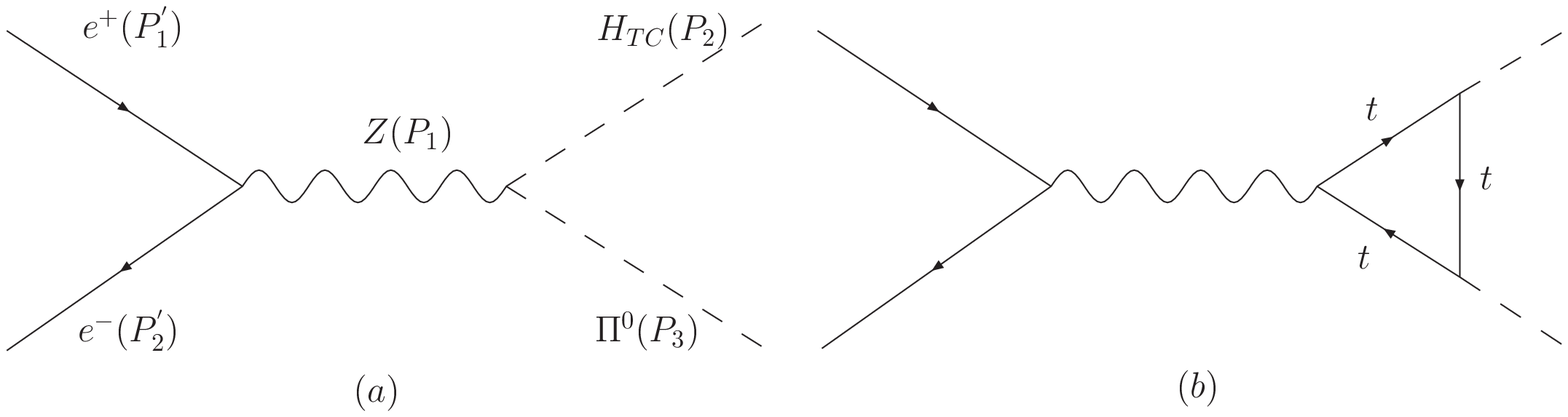} \figcaption{The Feynman
diagrams for $ e^ + e^ - \to H_{TC} \Pi ^0 $. } \label{ZHP}
\end{minipage}\\
\\[\intextsep]

With the NLO amplitude, we have obtained the NLO differential cross
section in the center-of-mass frame:

\begin{eqnarray}
\frac{d\sigma _{NLO} }{d\Omega } = \frac{1}{2S}\frac{\vert \vec{P_2}
\vert }{16\pi ^2\sqrt S }\frac{1}{4}\sum\limits_{spins} {(\vert
M^0\vert } ^2 + 2Re[(M^1)^\dag M^0] + \vert M^1\vert ^2).
\end{eqnarray}\\

Integration over the solid angle, we have the total cross section.\\

It is noticed that in the process $ e^ + e^ - \to H_{TC} \Pi ^0 $,
the ratio of $\vert M^1\vert ^2/\vert M^0\vert ^2 \geq 0.20$ at most
of the parameter spaces and cannot be thrown away.\\

\subsection{ For $ e^ + e^ - \to \Pi^{+} \Pi ^- $}\vspace{5mm}

The Feynman diagrams responsible for this process are shown in Fig.
\ref{ZPP}.\\

The tree level amplitude of $ e^ + e^ - \to \Pi^{+} \Pi ^- $ is:
\begin{eqnarray}
\begin{array}{l}
 M^{0'} = M_\gamma ^{0'} + M_Z^{0'} ,\\
~~\\
 M_\gamma ^{0'} = - i\frac{\displaystyle g^2s_w (1 - 2s_w^2 )}{\displaystyle c_w }\frac{\displaystyle 1}{\displaystyle P_1^2 }\bar
{v}(P_1^{'} )(\not {P}_3 - \not {P}_2 )u(P_2^{'} ) ,\\
~~\\
 M_Z^{0'} = i\frac{\displaystyle g^2(1 - 2s_w^2 )}{\displaystyle c_w^2 }\frac{\displaystyle 1}{\displaystyle P_1^2 - M_Z^2 }\bar
{v}(P_1^{'} )(\not {P}_3 - \not {P}_2 )[ - \frac{\displaystyle
1}{\displaystyle 4}(1 - \gamma _5 ) + s_w^2
]u(P_2^{'} ) .\\
 \end{array}
\end{eqnarray}\\

The loop-induced amplitude can be written in the form:\\
\begin{eqnarray}
 M^{loop'} = M_z^{loop} + M_\gamma ^{loop},
\end{eqnarray}
where the subscripts "Z" and "$\gamma$" correspond to the diagrams
where $Z$ boson or photon is exchanged. By the Lorentz structure of
the coupling, one can immediately show
$$ M_\gamma ^{loop} = 0, $$
and then
\begin{equation}
 M^{loop'} = M_z^{loop} = \frac{\displaystyle g^2}{\displaystyle 2cw^2}N\bar {\displaystyle v}(P_1^{'} )[f_2^{'} \not
{P}_2 + f_3^{' }\not {P}_3 ][ - \frac{\displaystyle 1}{\displaystyle
4}(1 - \gamma _5 ) + s_w^2 ]u(P_2^{'}
)\frac{\displaystyle 1}{\displaystyle P_1^2 - M_z^2 }. \\
\end{equation}\\

The explicit expressions of relevant form factors ($f_2^{'}$,
$f_3^{'}$) are presented in Appendix A.\\
\\[\intextsep]
\begin{minipage}{\textwidth}
\centering
\includegraphics[height=3cm,angle=0]{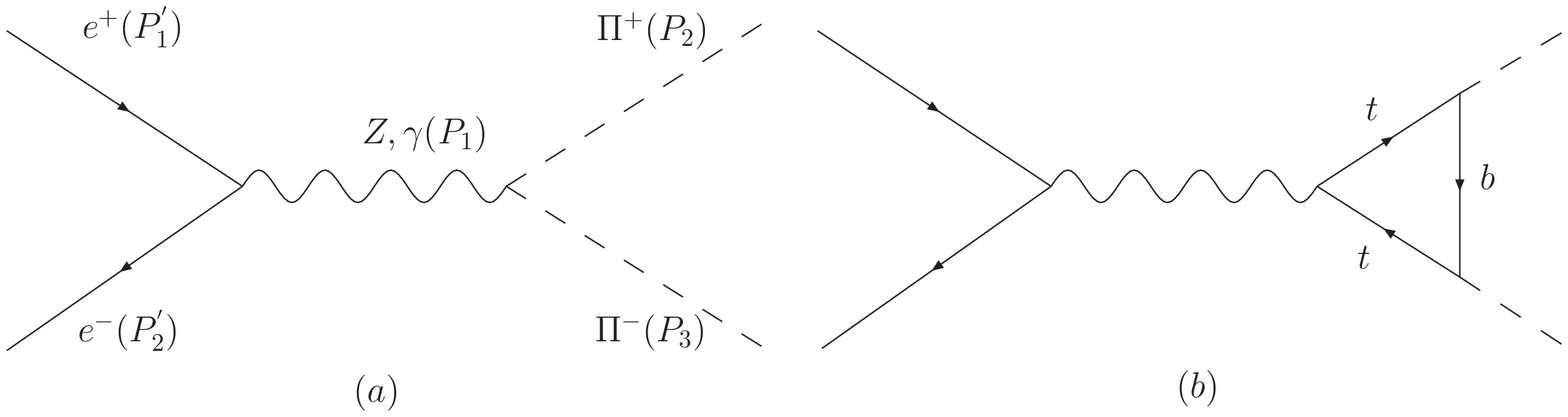} \figcaption{The Feynman
diagrams for $ e^ + e^ - \to \Pi^+ \Pi ^- $. } \label{ZPP}
\end{minipage}\\

\section{The numerical results} \hspace{0mm}\vspace{2mm}\\

To obtain numerical results of the cross sections, we adopt the
input parameters $M_Z=91.188$ GeV, $s_w ^2=0.23$ and $v_\pi=100$
GeV. In our calculations the mass of top Higgs takes two different
values: $M_H=200,\; 300$ GeV \cite{Hill:2002ap}. The electromagnetic
fine-structure constant $\alpha$ at the concerned energy scale is
calculated by the renormalization group equation (RGE) with the
boundary value $\alpha^{-1}=137.04$.  Generally in the TC2 model,
the mass of top pions is supposed to be around 200 GeV, for a
phenomenological study, we let the mass vary within a narrow range
of $150\sim300$ GeV. Following the general discussion about the
choice of center-of-mass energy for ILC, in our calculation it is
set as $\sqrt S$ = 500 GeV \cite{ILC}. The numerical results of the
cross sections are shown in Fig. \ref{mh200}
and Fig. \ref{mh300}\\
\\[\intextsep]
\begin{minipage}[c]{0.5\textwidth}
\centering
\includegraphics[width=2.5in]{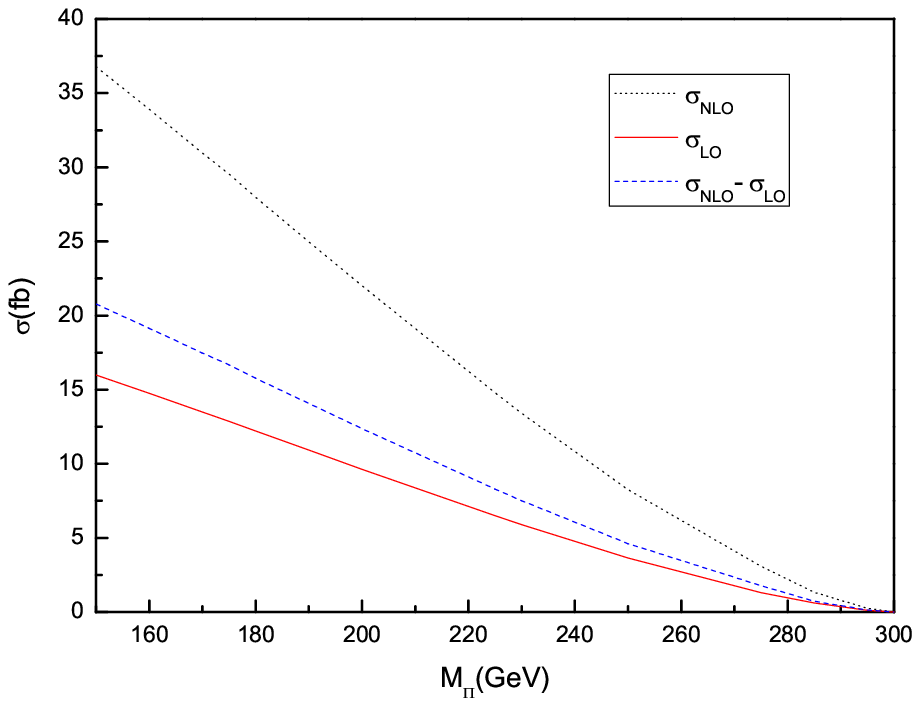}
\end{minipage}%
\begin{minipage}[c]{0.5\textwidth}
\centering
\includegraphics[width=2.5in]{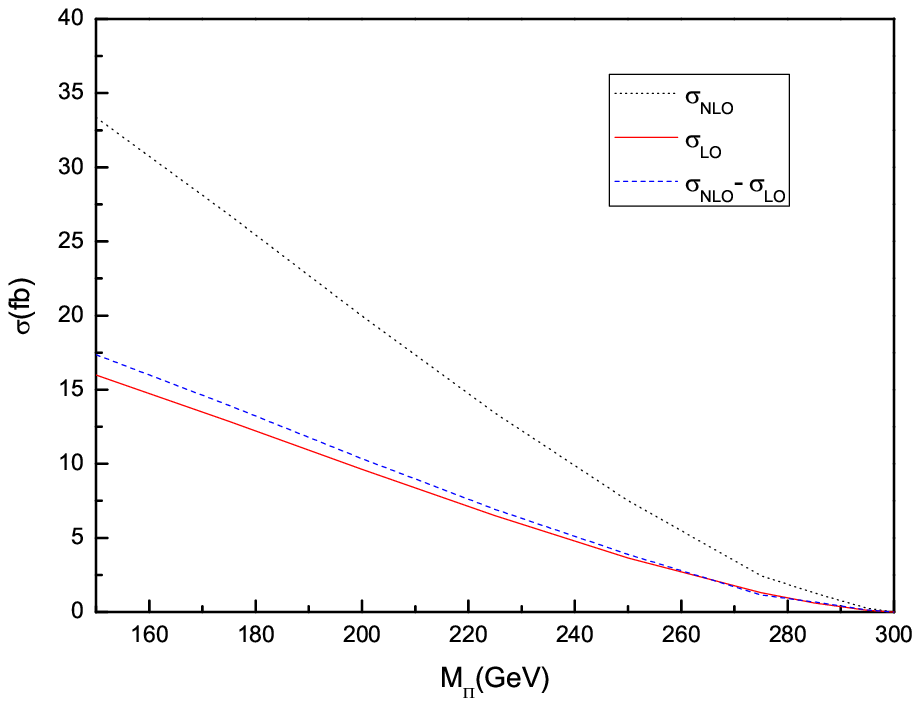}
\end{minipage}\\
\begin{minipage}{\textwidth}
\figcaption{Dependence of the cross section of  $ e^ + e^ - \to
H_{TC} \Pi ^0 $ on top-pion mass $M_\Pi$ (150$\sim$300 GeV) for
$M_H$ =200 GeV , $\epsilon_t=0.03$ (left) and $\epsilon_t=0.1$
(right) respectively.}\label{mh200}
\end{minipage}\\
\\[\intextsep]
\\[\intextsep]
\begin{minipage}[c]{0.5\textwidth}
\centering
\includegraphics[width=2.5in]{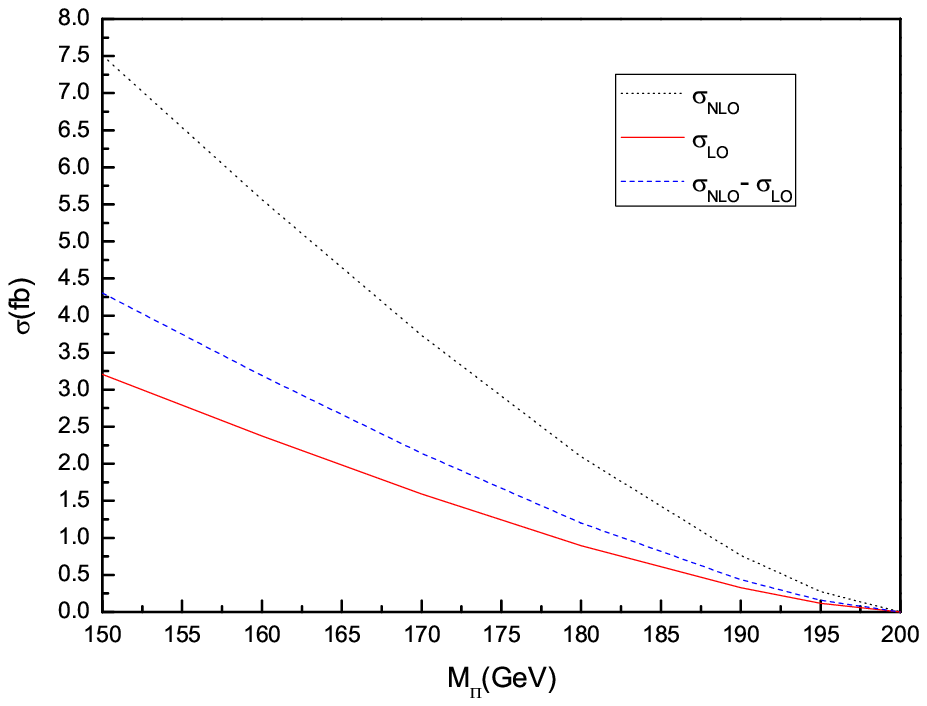}
\end{minipage}%
\begin{minipage}[c]{0.5\textwidth}
\centering
\includegraphics[width=2.5in]{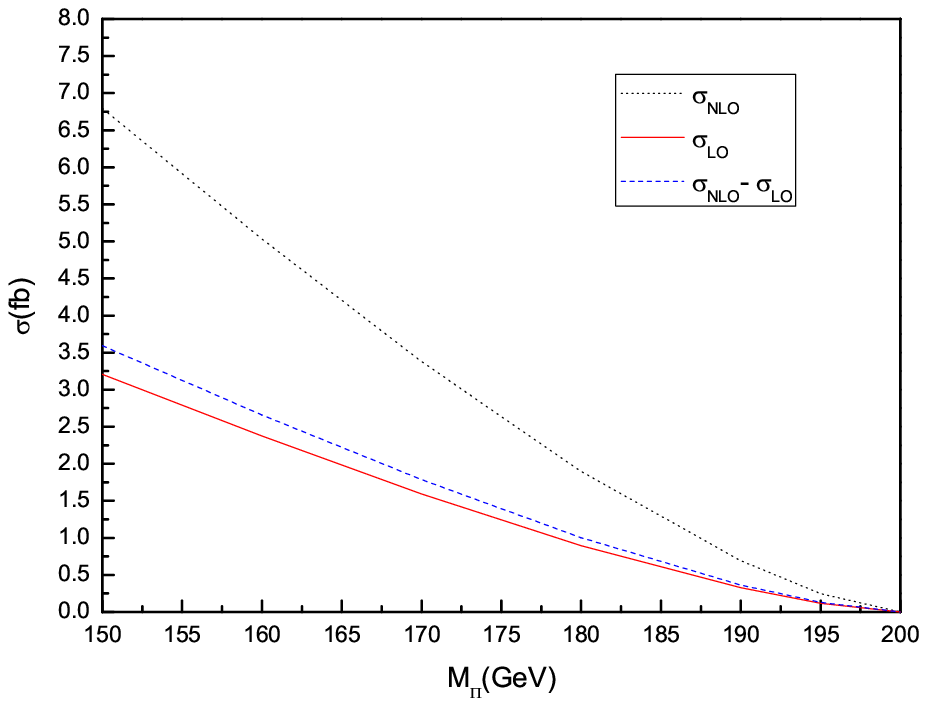}
\end{minipage}\\
\begin{minipage}{\textwidth}
\figcaption{Dependence of the cross section of  $ e^ + e^ - \to
H_{TC} \Pi ^0 $ on top-pion mass $M_\Pi$ (150$\sim$200 GeV) for
$M_H$ =300 GeV , $\epsilon_t=0.03$ (left) and $\epsilon_t=0.1$
(right) respectively.}\label{mh300}
\end{minipage}\\
\\[\intextsep]

The plots show that the cross section decreases with $M_\Pi$  and
$M_{H_{TC}}$  and it is also noticed that as $\epsilon_t$ takes a
value of 0.03 the cross section drops slightly faster than that for
the case of $\epsilon_t=0.1$, namely it is not very sensitive to the
value of $\epsilon$ which tells how the different sources contribute
to the top quark mass. In general, the production rate is at the
level of a few fb. Through the figures we also can observe that the
loop-induced correction $\delta \equiv (\sigma _{NLO} - \sigma _{LO}
)/\sigma _{LO} $ exceeds 100\%  at most parameter spaces and even
exceeds 130\% for extreme situations. The dependence of $\delta$ on
the input parameters can be seen in Table 1 of Appendix
B.\\

The corresponding differential cross sections (DCS) are shown in
Fig. \ref{dcsph} where the parameters are  explicitly listed. From
the figures, we can see that the DCS are symmetric with respect to
$\pi/2$ and DCS decreases as $M_\Pi$ increases.
\\[\intextsep]
\begin{minipage}[c]{0.5\textwidth}
\centering
\includegraphics[width=3.5in]{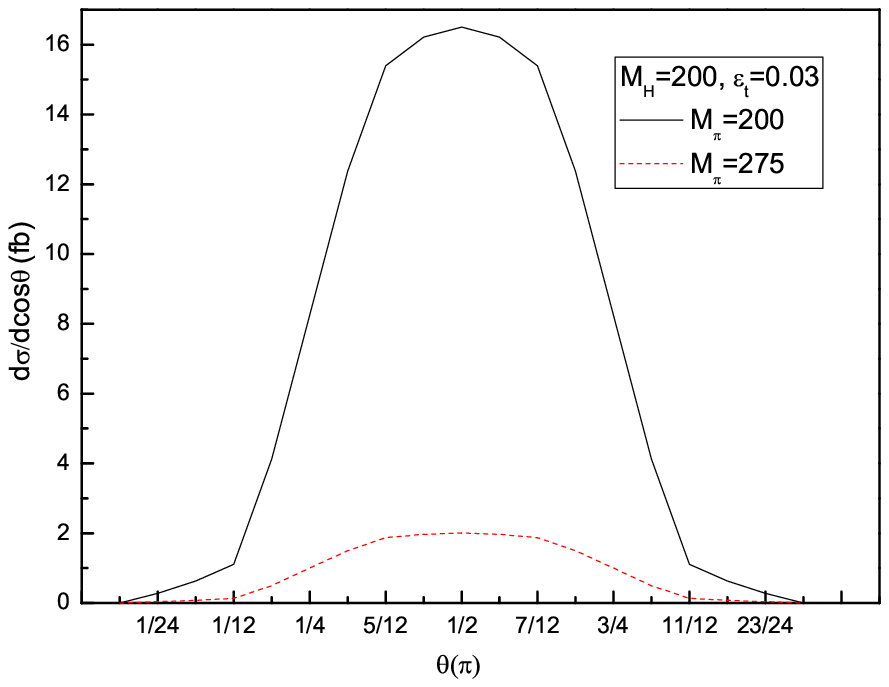}
\end{minipage}%
\begin{minipage}[c]{0.5\textwidth}
\centering
\includegraphics[width=3.5in]{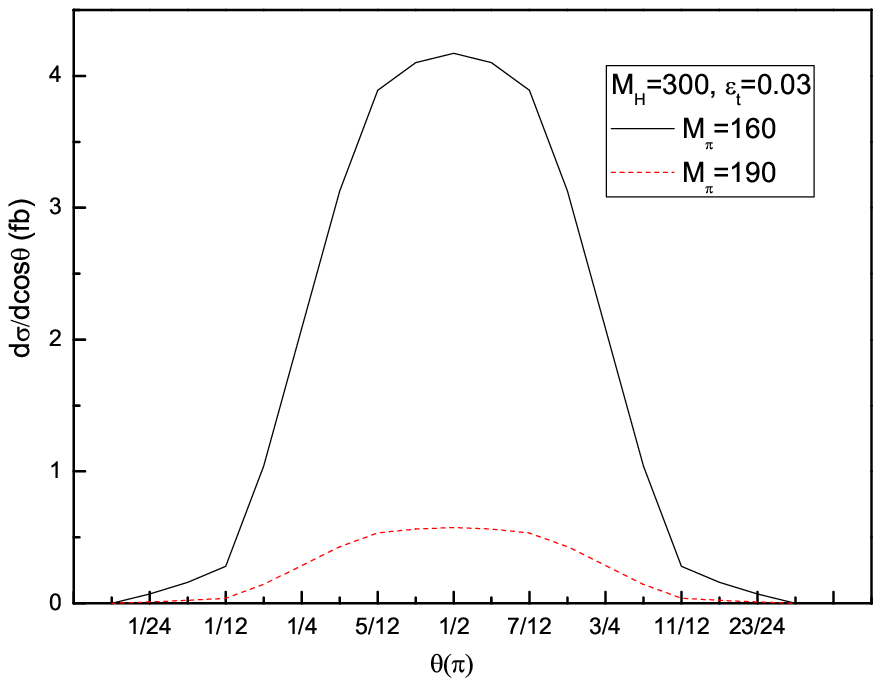}
\end{minipage}\\
\begin{minipage}{\textwidth}
\figcaption{The dependence of the differential cross section of  $
e^ + e^ - \to \ H_{TC} \Pi ^0$ on $\theta$.}\label{dcsph}
\end{minipage}\\
\\[\intextsep]

If the ILC energy is upgraded up to 1 TeV \cite{ILC} (i.e.
$\sqrt{s}=1$ TeV), the NLO correction to the process will further
increase as shown in Fig. \ref{ph1000003}:
\\[\intextsep]
\begin{minipage}[c]{\textwidth}
\centering
\includegraphics[width=4.6in]{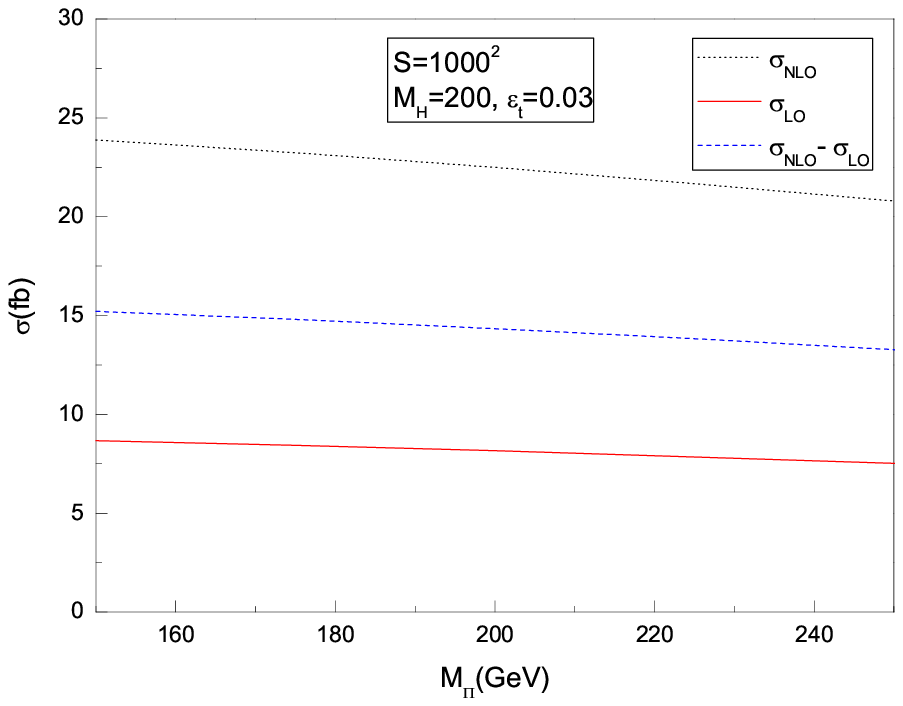}
\end{minipage}
\begin{minipage}{\textwidth}
\figcaption{The cross section of  $ e^ + e^ - \to H_{TC} \Pi ^0 $
with $\sqrt{s}=1$ TeV.}\label{ph1000003}
\end{minipage}\\
\\[\intextsep]

We can see that when $\sqrt{s}=1$ TeV, the NLO correction is even
more important and $\delta$ exceeds that for $\sqrt{s}=500$ GeV.\\

The numerical results for  $ e^ + e^ - \to \Pi^+ \Pi ^- $ are shown
in Fig. \ref{mp}.
\\[\intextsep]
\begin{minipage}[c]{0.5\textwidth}
\centering
\includegraphics[width=3in]{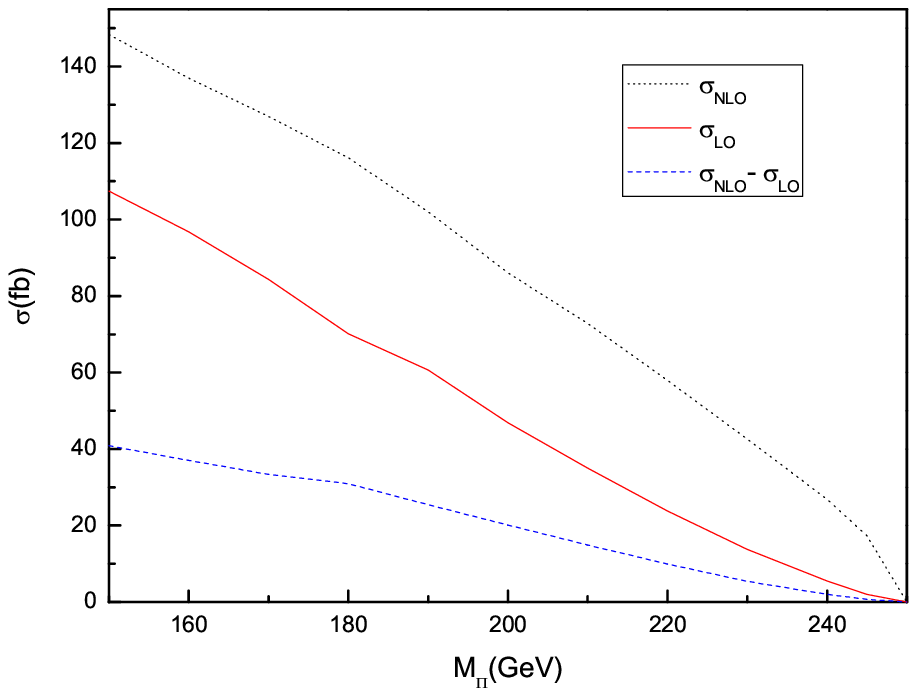}
\end{minipage}%
\begin{minipage}[c]{0.5\textwidth}
\centering
\includegraphics[width=3in]{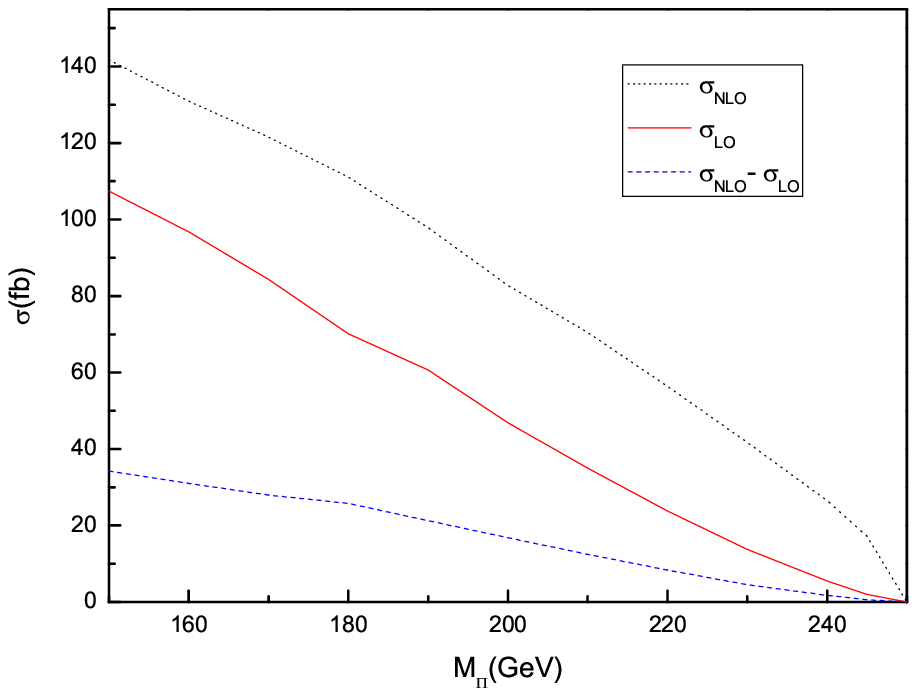}
\end{minipage}\\
\begin{minipage}{\textwidth}
\figcaption{Dependence of The cross section of  $ e^ + e^ - \to
\Pi^+ \Pi ^- $ on top-pion mass $M_\Pi$ (150$\sim$250 GeV) for
$\epsilon_t=0.03$ (left) and $\epsilon_t=0.1$ (right)
respectively.}\label{mp}
\end{minipage}\\
\\[\intextsep]

The NLO corrections for $\sqrt{s}=1$ TeV are shown in Fig.
\ref{pp1000003}:
\\[\intextsep]
\begin{minipage}[c]{\textwidth}
\centering
\includegraphics[width=4.6in]{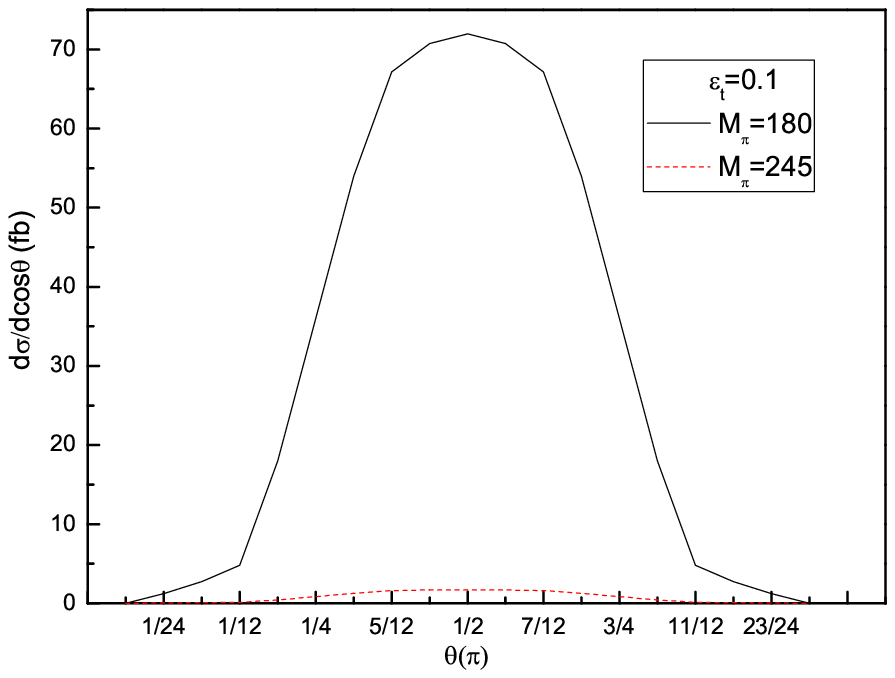}
\end{minipage}
\begin{minipage}{\textwidth}
\figcaption{Dependence of the differential cross section of  $ e^ +
e^ - \to \Pi^+ \Pi ^- $ on $\theta$ with $\sqrt{s}=500$
GeV.}\label{Gpp01}
\end{minipage}\\
\\[\intextsep]
\\[\intextsep]
\begin{minipage}[c]{\textwidth}
\centering
\includegraphics[width=4.6in]{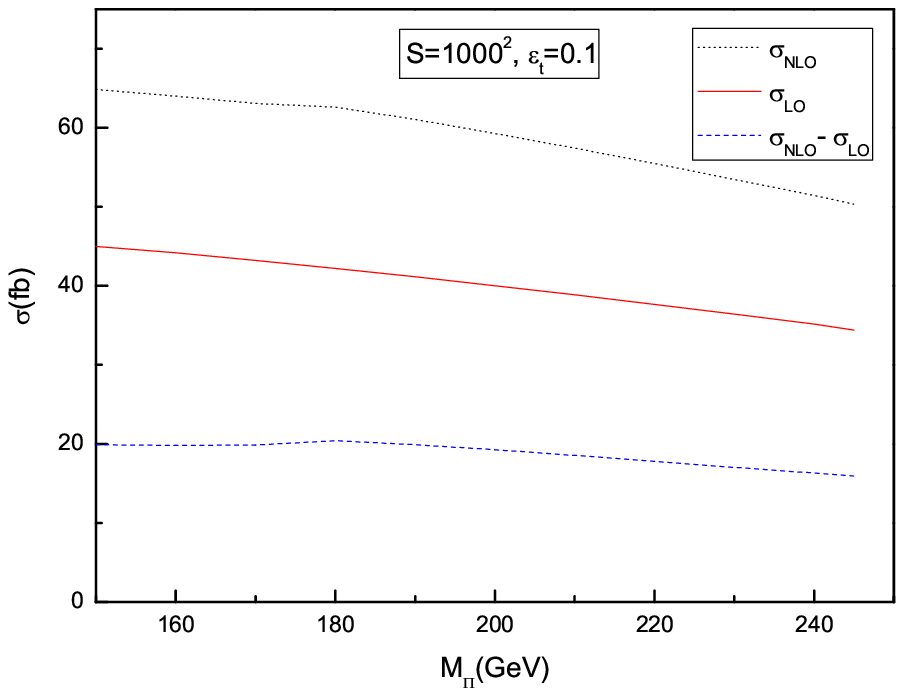}
\end{minipage}
\begin{minipage}{\textwidth}
\figcaption{The cross section of  $ e^ + e^ - \to \Pi^+ \Pi ^- $
with $\sqrt{s}=1$ TeV.}\label{pp1000003}
\end{minipage}\\
\\[\intextsep]

The dependence of the relative correction $\delta$ on the
input parameters is presented in the Table 2 of Appendix B.\\

Our results indicate that the NLO corrections to $ e^ + e^ - \to
\Pi^+ \Pi ^- $ are not as significant as to the $ e^ + e^ - \to
H_{TC} \Pi ^0$, this is because there is an extra  contribution from
another tree diagram where a virtual photon serves as the
intermediate state, thus
the loop contributions are relatively smaller than the total tree contributions.\\

\section{THE CONCLUSIONS AND OUTLOOK} \hspace{0mm}\vspace{2mm}\\

Through the Tables in Appendix B, one can see that with yearly
designed luminosity about 500 $fb^{-1}$ at ILC \cite{ILC}, if the
detection efficiency can be 20\%, more than several $10^3$
$H_{TC}\Pi^0$ and about $10^4$ $\Pi^+\Pi^-$ signals can be expected
at ILC as long as the mass of relevant particles and corresponding
parameters reside in a reasonable region,  and if the luminosity can
reach 1000 $fb^{-1}$, the amount of signals would be doubled and
detection of such TC2 particles $\Pi^{\pm,0}$ and $H_{TC}$
would be very optimistic.\\

Our calculations indicate that the NLO contributions are important
for the two concerned processes which may be crucial for detecting
the TC2 model. The reason for larger NLO correction may be twofold.
Firstly, in the loop, the intermediate states are top
quark-antiquark whose mass in the TC2 theory is determined by two
sources and expressed as \cite{Leibovich:2001ev}: $M_t=-
\frac{\displaystyle 1}{\displaystyle \sqrt 2 }(Y_t f_\pi + \epsilon
_t v_T )$, and its  TC Yukawa coupling $Y_t$ is high and causes an
enhancement. Secondly, the extra color factor $N_c=3$ in the loop
will further increase the loop-induced amplitude.\\

Because of the relative high one-loop contribution, one may
naturally ask if two-loops contributions are necessary. If the two
arguments listed above are the only reasons, we may expect  that the
two-loop contribution would not exceed the one-loop contribution.
\\

The decay modes of $H_{TC}$, $\Pi ^ \pm, \Pi ^
0$\cite{Wang:2004,Yue:2000fe} and a comparison with that in other
models beyond the SM \cite{Djouadi:1996pj} at $e^+e^-$ linear
colliders have been discussed in Ref. \cite{Wang:2005ixa}. We do not
intend to discuss these topics in this work, even though they are
crucially important for observation, and will come back to it in our
later work.
\\

The advantage of analyzing such processes at the ILC is obvious that
the hadronic background is very suppressed and the amount of signals
may be practically observable. The calculation of the production at
the $e^+e^-$ collision is relatively simple compared to the case for
hadron colliders because there is no QCD correction and moreover,
there does not exist the complicated infrared divergence which needs
to be properly dealt with. By contrast, the situation would be
deteriorated at the hadron colliders such as LHC, however, on other
aspect, the production probability of the new physics particles at
LHC may be much larger, so that the disadvantage caused by
background contamination may be compensated. But definitely, one
needs to consider the production rates of $H_{TC}$, $\Pi ^ \pm, \Pi
^ 0$ at LHC up to NLO and it would be our next work.\\

It is worth of noticing that, one may conjecture that a $H_{TC}$
pair or a  $\Pi^0$ pair may also be produced at the one-loop level,
but the results show that their contributions equal to zero due to
an obvious symmetry constraints. \\

Our conclusion is that if the Higgs boson were not observed at LHC,
the technicolor model would be favorable because it provides a
dynamical symmetry breaking mechanism. Then one needs to look for
evidence for existence or validity of the model, so  detection of
production of some specific particles which carry characteristics of
the model would be a direct trace. We calculate the production rates
of $ e^ + e^ - \to \Pi^+ \Pi ^- $ and $ e^ + e^ - \to H_{TC} \Pi ^0$
at ILC to NLO, supposing its CM energy to be 500 GeV and find that
the rates are sizable to be observed for a low background machine.
In the calculations, we also notice that the NLO contributions for
both modes are high compared to that of LO and then briefly analyze
the reason. Therefore we indicate that to compare the theoretical
prediction with data, one needs to carry out the calculation to NLO,
moreover, our simple analysis may imply the NNLO should be smaller
and less significant.

\newpage

$\Large{\mathbf{APPENDIX}}$\\

\appendix
\section{The explicit expressions of the form factors}\hspace{0mm}\vspace{2mm}\\

The explicit expressions of the form factors $f_i$ used in the paper
can be written as:\\

\begin{eqnarray}
f_2 = - \frac{1}{4\pi ^2}[\frac{M_t \tan \beta }{v}(1 - \epsilon
_t)]^2 [B_1 + (C_1 + C_{11} ) (S + M_\Pi ^2 - M_H^2 ) + (C_2 +
2C_{12} )M_\Pi ^2 - C_0 M_t^2 + B_0],
\end{eqnarray}

\begin{eqnarray}
\begin{array}{l}
 f_3 =  - \frac{\displaystyle 1}{\displaystyle 4\pi ^2}[\frac{\displaystyle M_t \tan \beta }
{\displaystyle v}(1 - \epsilon _t )]^2 [B_1 + C_1 (M_\Pi ^2 - M_H^2 ) + C_{11} (S + M_\Pi ^2 - M_H^2 ) + (C_2
+ 3C_{12} )M_\Pi ^2 \\
~~~~~~\\ ~~~~~~- 4C_0 M_t^2 + 2C_{00} + 2C_{22} M_\Pi ^2 + C_{12} (S - M_H^2 )].\\
 \end{array}\end{eqnarray}

\begin{eqnarray}
C_{ij}=C_{ij}(P_1^2,(-P_2)^2,(P_1-P_2)^2,M_t^2,M_t^2,M_t^2),
\end{eqnarray}

\begin{eqnarray}
B_0=B_0((-P_2)^2,M_t^2,M_t^2).
\end{eqnarray}

\begin{eqnarray}
f_2^{'} = i \frac{1}{4\pi ^2}[\frac{M_t \tan \beta }{v}(1 - \epsilon
_t )]^2 [B_0^{'} + B_1^{'} + S(C_1^{'} + C_{11}^{'} ) + M_\Pi ^2
(C_2^{'} + 2C_{12}^{'} ) + M_t^2 C_0^{'} ],
\end{eqnarray}

\begin{eqnarray}
\begin{array}{l}
 f_3^{'} =  i \frac{\displaystyle 1}{\displaystyle 4\pi ^2}[\frac{\displaystyle M_t
\tan \beta }{\displaystyle v}(1 - \epsilon _t )]^2 [B_1^{'} + S(C_{11}^{'} + C_{12}^{'} ) +
M_\Pi ^2 (C_2^{'} + 2C_{12}^{'} + 2C_{22}^{'} + M_t^2 C_0^{'})~~~~~~\\~\\ ~~~~~~+ 2C_{00}^{'}
 - M_t^2 C_0^{'} ]. \\
 \end{array}\end{eqnarray}

\begin{eqnarray}
C_{ij}^{'}=C_{ij}^{'}(P_1^2,(-P_2)^2,(P_1-P_2)^2,M_t^2,M_t^2,M_b^2),
\end{eqnarray}

\begin{eqnarray}
B_0^{'}=B_0^{'}((-P_2)^2,M_t^2,M_b^2).
\end{eqnarray}\\

Here $B_i$, $C_{ij}$ are two-point and three-point scalar integrals.
$P$ represents the momentum of relevant particle. The explicit
Lorentz decompositions
for the lowest order integrals take the forms given in Ref. \cite{Denner:1991kt}\\

\newpage

\section{The ratio of $(\sigma _{NLO} - \sigma _{LO} )/\sigma _{LO}
$}\hspace{0mm}\vspace{2mm}
\begin{table}[!h]
\tabcolsep 0pt \caption{Dependence of $\delta$ on the parameter in $
e^ + e^ - \to H_{TC} \Pi ^0 $} \vspace*{-12pt}
\begin{center}
\def\temptablewidth{0.5\textwidth}
{\rule{\temptablewidth}{1pt}}
\begin{tabular*}{\temptablewidth}{@{\extracolsep{\fill}}ccccccc}
$\epsilon_t $ &$M_H$ & $M_\Pi$ &$\sigma_{NLO}$ &$\sigma_{LO}$ &$\delta$ \\
\hline
      0.1 & 200 & 150 & 33.3611  & 15.9941   & 108.584\%  \\
          &     & 175 & 26.7926  & 12.8525   & 108.462\%  \\
          &     & 200 & 19.9713  & ~9.6295   & 107.397\%  \\
          &     & 225 & 13.4233  & ~6.4926   & 106.749\%  \\
          &     & 250 & ~7.5164  & ~3.6369   & 106.669\%  \\
          &     & 275 & ~2.4542  & ~1.3117   & ~87.098\%  \\
          &     & 285 & ~1.2881  & ~0.6130   & 110.139\%  \\
          &     & 295 & ~0.2432  & ~0.1184   & 105.321\%  \\
\hline
      0.03& 200 & 150 & 36.7804  & 15.9941   & 129.963\%  \\
          &     & 175 & 29.5348  & 12.8525   & 129.798\%  \\
          &     & 200 & 22.0070  & ~9.6295   & 128.537\%  \\
          &     & 225 & 14.7877  & ~6.4926   & 127.764\%  \\
          &     & 250 & ~8.2565  & ~3.6369   & 127.018\%  \\
          &     & 275 & ~2.6729  & ~1.3117   & 103.772\%  \\
          &     & 285 & ~1.4235  & ~0.6130   & 132.234\%  \\
          &     & 295 & ~0.2678  & ~0.1184   & 126.100\%  \\
\hline \hline
      0.1 & 300 & 150 & ~6.7968 & ~3.2051  & 112.061\%  \\
          &     & 160 & ~5.0331 & ~2.3744  & 111.975\%  \\
          &     & 170 & ~3.3789 & ~1.5940  & 111.980\%  \\
          &     & 180 & ~1.8980 & ~0.8953  & 111.989\%  \\
          &     & 190 & ~0.6914 & ~0.3261  & 112.003\%  \\
          &     & 195 & ~0.2480 & ~0.1170  & 112.012\%  \\
\hline
      0.03& 300 & 150 & ~7.5093 & ~3.2051  & 134.290\%  \\
          &     & 160 & ~5.5606 & ~2.3744  & 134.191\%  \\
          &     & 170 & ~3.7331 & ~1.5940  & 134.197\%  \\
          &     & 180 & ~2.0969 & ~0.8953  & 134.209\%  \\
          &     & 190 & ~0.7639 & ~0.3261  & 134.228\%  \\
          &     & 195 & ~0.2740 & ~0.1170  & 134.240\%  \\

       \end{tabular*}
       {\rule{\temptablewidth}{1pt}}
       \end{center}
       \end{table}\\

\hspace{0mm}\vspace{2mm}
\begin{table}[!t]
\tabcolsep 0pt \caption{Dependence of $\delta$ on the parameter in $
e^ + e^ - \to \Pi^+ \Pi ^- $} \vspace*{-12pt}
\begin{center}
\def\temptablewidth{0.5\textwidth}
{\rule{\temptablewidth}{1pt}}
\begin{tabular*}{\temptablewidth}{@{\extracolsep{\fill}}ccccccc}
$\epsilon $ &$M_\Pi$ &$\sigma_{NLO}$ &$\sigma_{LO}$ &$\delta$ \\
\hline
      0.1 & 150 & 141.7408 & 107.4589  & 31.902\%  \\
          & 160 & 127.8301 & ~96.7722  & 32.093\%  \\
          & 170 & 112.2826 & ~84.3145  & 33.171\%  \\
          & 180 & ~95.9705 & ~70.1459  & 36.815\%  \\
          & 190 & ~81.9590 & ~60.6831  & 35.060\%  \\
          & 200 & ~63.6793 & ~46.8669  & 35.872\%  \\
          & 210 & ~47.4190 & ~34.9830  & 35.548\%  \\
          & 220 & ~32.1575 & ~23.8273  & 34.960\%  \\
          & 230 & ~18.3033 & ~13.7898  & 32.731\%  \\
          & 240 & ~~7.1690 & ~~5.4693  & 31.076\%  \\
          & 245 & ~~2.5796 & ~~1.9700  & 30.942\%  \\

\hline
      0.03 & 150 & 148.3453  & 107.4589 & 38.048\%  \\
           & 160 & 133.8350  & ~96.7723 & 38.298\%  \\
           & 170 & 117.7175  & ~84.3145 & 39.617\%  \\
           & 180 & 101.0672  & ~70.1459 & 44.081\%  \\
           & 190 & ~86.1166  & ~60.6831 & 41.911\%  \\
           & 200 & ~66.9610  & ~46.8669 & 42.875\%  \\
           & 210 & ~49.8442  & ~34.9830 & 42.481\%  \\
           & 220 & ~33.7806  & ~23.8273 & 41.772\%  \\
           & 230 & ~19.1579  & ~13.7898 & 38.928\%  \\
           & 240 & ~~7.4997  & ~~5.4693 & 37.123\%  \\
           & 245 & ~~2.6982  & ~~1.9700 & 36.962\%  \\

       \end{tabular*}
       {\rule{\temptablewidth}{1pt}}
       \end{center}
       \end{table}\\

\end{document}